\documentclass[aps,prd,showpacs,preprintnumbers,twocolumn]{revtex4}
\usepackage{graphicx}
\usepackage{epsfig}
\usepackage{amsmath}
\usepackage{subfigure}
\usepackage{ulem}
\usepackage{color}
\usepackage{latexsym} 
\usepackage{amssymb}  
\usepackage{amsfonts} 
\usepackage{amsbsy}   
\usepackage{multirow} 
\usepackage{pstricks}
\usepackage{slashed}
\usepackage{dcolumn}
\usepackage{xparse}
\DeclareMathOperator{\sech}{sech}

\newcommand{\al}{\alpha}
\newcommand{\be}{\beta}
\newcommand{\ga}{\gamma}
\newcommand{\Ga}{\Gamma}

\newcommand{\De}{\Delta}

\newcommand{\eps}{\epsilon}

\newcommand{\La}{\Lambda}

\renewcommand{\th}{\theta}   

\newcommand{\Th}{\Theta}

\newcommand{\beq}{\begin{equation}}
\newcommand{\eeq}{\end{equation}}
\newcommand{\ba}{\begin{array}}
\newcommand{\bea}{\begin{eqnarray}}
\newcommand{\ea}{\end{array}}
\newcommand{\eea}{\end{eqnarray}}
\newcommand{\bi}{\begin{itemize}}    
\newcommand{\ei}{\end{itemize}}
\newcommand{\ben}{\begin{enumerate}} 
\newcommand{\een}{\end{enumerate}}
\newcommand{\bc}{\begin{center}}
\newcommand{\ec}{\end{center}}
\newcommand{\bl}{\begin{flushleft}}
\newcommand{\el}{\end{flushleft}}
\newcommand{\br}{\begin{flushright}}
\newcommand{\er}{\end{flushright}}

\newcommand{\nn}{\nonumber \\}
\newcommand\eqn[1]{(\ref{#1})}      
\newcommand\Eqn[1]{Eq.~(\ref{#1})}  
\newcommand{\mr}{\mathrm}
\newcommand{\mb}{\mathbf}

\newcommand{\mc}{\mathcal}
\newcommand{\mi}{\mathrm{i}}
\newcommand{\me}{\mathrm{e}}
\newcommand{\dif}{\mathrm{d}}
\newcommand{\p}{\partial}

\newcommand{\Tr}{\hbox{Tr}}

\renewcommand{\Re}{{\mathrm{Re}}\,}
\renewcommand{\Im}{{\mathrm{Im}}\,}

\newcommand{\<}{\langle}
\renewcommand{\>}{\rangle}   
\renewcommand{\l}{\left}
\renewcommand{\r}{\right}

\newcommand\comment[1]{ \hbox{[{\it Comment suppressed here.}\/]} }
\newcommand\hide[1]{}
\newcommand{\skipover}[1]{}
\newcommand{\kp}{k_{\shortparallel}}
\newcommand{\qp}{q_{\shortparallel}}
\newcommand{\gp}{g_{\shortparallel}}
\newcommand{\kt}{k_{\perp}}
\newcommand{\qt}{q_{\perp}}
\newcommand{\gt}{g_{\perp}}
\newcommand{\hkp}{\hat{k}_{\shortparallel}}
\newcommand{\hqp}{\hat{q}_{\shortparallel}}

\newcommand{\hkt}{\hat{k}_{\perp}}
\newcommand{\hqt}{\hat{q}_{\perp}}

\newcommand{\epst}{\eps_{\perp}}
\newcommand{\tone}{t_{1}}
\newcommand{\ttwo}{t_{2}}
\begin{document}
\title{Zeta Function Regularization of Photon Polarization Tensor for a Magnetized Vacuum}
\author{Jingyi Chao$^{1}$}
\email{chaojy@mail.ihep.ac.cn}
\author{Lang Yu$^{1}$}
\email{yulang@mail.ihep.ac.cn}
\author{Mei Huang$^{1,2}$}
\email{huangm@mail.ihep.ac.cn}
\affiliation{$^{1}$ Institute of
High Energy Physics, Chinese Academy of Sciences, Beijing, China}
\affiliation{$^{2}$ Theoretical Physics Center for Science
Facilities, Chinese Academy of Sciences, Beijing, China}
\date{\today }
\bigskip
\date{\today}
\begin{abstract}
In this paper, we have developed a systematic technique to regularize double summations of Landau levels and analytically evaluated the photon vacuum polarization at an external magnetic field. The final results are described by Lerch transcendent $\Phi(z,s,v)$ or its $z$-derivation. We have found that the tensor of vacuum polarization is split into not only longitudinal and transverse parts but also another mixture component. We have obtained a complete expression of the magnetized photon vacuum polarization at any kinematic regime and any strength of magnetic field for the first time. In the weak $B$-fields, after canceling out a logarithmic counter term, all three scalar functions are limited to the usual photon polarization tensor  without turning on magnetic field. In the strong $B$-fields, the calculations under Lowest Landau Level approximation are only valid at the region $M^2\gg q_{\shortparallel}^2$, but not correct while $q_{\shortparallel}^2\gg M^2$, where, an imaginary part has been missed. It reminds us, a recalculation of the gap equation under a full consideration of all Landau Levels is necessary in the next future.
\end{abstract}
\maketitle
\section{Introduction}
\label{sec:intro}
It has been a long time ago that Schwinger studied the physical problems of fermions moving in a constant magnetic field \cite{schwinger1951gauge}. Working in Landau and symmetric gauges, he obtained the exact fermion propagator written in an integral representative. In 1990, Chodos \textit{et al.} demonstrated that the propagator is equivalent to a summation representative by decomposing over Landau poles \cite{chodos1990qed}. The summation of Landau levels clearly indicates that the related calculations of magnetized matter are more complicated than usual ones. Especially, if the infinite series are not convergent, it would be hard to extract the finite results by general regularization methods, such as Pauli-Villars, dimensional regularization, and so on.

Basically, Quantum Electrodynamics (QED) vacuum is modified after turning on an external electromagnetic field and then several striking phenomena are created, such as photon decay into an electron-positron pair via Schwinger mechanism \cite{schwinger1951gauge,tsai1974photon,baier2007pair}; vacuum birefringence of a photon \cite{meszaros1979vacuum,dittrich1998vacuum}; photon splitting and so on \cite{adler1971photon,adler1996photon}. As a fundamental information of magnetized vacuum, photon vacuum polarization tensor \cite{melrose1977polarization,calucci1994nonlogarithmic,hattori2013vacuum,karbstein2013photon,ishikawa2013numerical} is expressed in terms of a double summation of infinite series with respect to two Landau levels occupied by virtual charged particles. Therefore, even though the tensor structure of one-loop diagram is plain to carry out, the subsequent calculations are particularly complicate to approach. Before, most works were focusing on the strong filed limit where an assumption of Lowest Landau Level (LLL) has been applied \cite{calucci1994nonlogarithmic,gusynin1996dimensional}. However, it is not clear how solid of LLL approximation is and what kind of invalid physics would be caused. Without a complete description of vacuum polarization, it also limits us to explore other possible non-trivial phenomena in response to the external electromagnetic fields.

The regularization and renormalization procedures are essential ingredients of quantum field theory. Among these methods, a powerful technique named as zeta function regularization is always greatly expected when encountering infinite sums \cite{elizalde1994zeta,dunne2005heisenberg}. For instance, the thermal radiation \cite{hawking1977zeta} and Casimir effect \cite{elizalde2006uses} have been computed by the zeta function directly. After our investigation, we confirm that the zeta function technique can be utilized to address the photon vacuum polarization tensor $\Pi^{\mu\nu}$ in the presence of $B$, as well. Remarkably, we find that the Lerch transcendent \cite{erdelyi1953higher,jeffrey2007table}, belonging to zeta family, is very efficient to handle the double infinite series of Landau levels. After regularizing the divergent series, it gives us a finite value of $\Pi^{\mu\nu}$ despite of the double summations.

Recently, an extremely strong magnetic field of the strength of $10^{18-20}\textmd{G}$, equivalent to the order of $m_{\pi}^{2}$, has been realized in non-central heavy-ion collisions at the Relativistic Heavy Ion Collider (RHIC) or the Large Hadron Collider (LHC) \cite{skokov2009estimate,deng2012event}. It draws a lot of interests to explore the theoretical problems of Quantum Chromodynamics (QCD) vacuum and strong-interaction matter under hadron-scaled magnetic fields. Hence, the purposes of this paper are double folds. Firstly, we are going to develop a systematic technique to extract finite results after double summing infinite Landau levels in the existence of external magnetic fields. We obtain a complete expression of the magnetized photon vacuum polarization at any kinematic regime and any magnitude of $B$. Secondly, after a full theoretical formula has been established, we wish our work can be served as a standard routine and toolbox to involve in future researches in QCD$\times$QED physics.

The paper is organized as follows. We demonstrate the procedure on how to derive the basic form of $\Pi^{\mu\nu}$ at external $B$ in section \ref{sec:vac}. Dimension regularization in longitudinal space also is explained here. In section \ref{sec:zeta}, we present how powerful of the zeta regularization method in extracting finite physical results from infinite series. We examine the limiting behaviors of $\Pi^{\mu\nu}$ and physical explanations at different kinematics regimes in section \ref{sec:res}. We end up with the summary and future applications in section \ref{sec:con}. In appendices, we provide supplemental materials of Lerch transcendent and some useful mathematic formula.
\section{Vacuum Polarizations}
\label{sec:vac}
The decomposed fermion propagator $\slashed{S}(k)$ is written as \cite{gusynin1996dimensional}:
\beq
\slashed{S}(k)=i\exp\l(2\hkt^{2}\r)\sum_{n=0}^{\infty}\frac{(-1)^{n}\slashed{D}_{n}(eB,k)}{k_{0}^{2}-k_{3}^{2}-M^{2}-2neB},
\eeq
where
\bea
&&\slashed{D}_{n}(eB,k)=2(\slashed{k}_{\shortparallel}+M)\mathcal{O}^{-}L_{n}(-4\hkt^{2})
\nn
&&-2(\slashed{k}_{\shortparallel}+M)\mc{O}^{+}L_{n-1}(-4\hkt^{2})-4\slashed{k}_{\perp}L_{n-1}^{1}(-4\hkt^{2})
\eea
with $\hkt=\kt^{2}/(2eB)$. $L_{n}^{a}(\xi)$ are the generalized Laguerre polynomials, where $L_{n-i}^{a}(\xi)=0$ if $n<i$. $\mc{O}^{\pm}=(1\pm\mi\ga^{1}\ga^{2})/2$ are the projecting operators corresponding to the spin state of charged particle paralleling or anti-paralleling to the direction of external field $B$. They satisfy following commutation relations: $\mc{O}^{\pm}\ga^{\mu}\mc{O}^{\pm}=\mc{O}^{\pm}\ga_{\shortparallel}$ and $\mc{O}^{\pm}\ga^{\mu}\mc{O}^{\mp}=\mc{O}^{\pm}\ga_{\perp}$. Here, the metric convention $g^{\mu\nu}$ is decomposed into two orthogonal subspaces $g^{\mu\nu}_{\shortparallel}=\mathrm{diag}(1,0,0,-1)$ and $g^{\mu\nu}_{\perp}=\mathrm{diag}(0,-1,-1,0)$. Similar decompositions are adopted for four dimensional momentum $k^{\mu}=k^{\mu}_{\shortparallel}+k^{\mu}_{\perp},~k^{\mu}_{\shortparallel}=(k^{0},0,0,k^{3}),~k^{\mu}_{\perp}=(0,k^{1},k^{2},0)$ and Dirac matrices $\ga^{\mu}_{\shortparallel}=(\ga^{0},0,0,\ga^{3}),~\ga^{\mu}_{\perp}=(0,\ga^{1},\ga^{2},0)$.

To calculate $\Pi^{\mu\nu}=-\mi e^2\Tr[\slashed{S}(k)\ga^{\mu}\slashed{S}(p)\ga^{\nu}]$, where $p=k+q$ and $e$ is the electrical charge. The general Feynman parameter for the denominator factor has been introduced \cite{peskin1995introduction}. Besides, Schwinger observed that all propagators may be rewritten as Gaussian integrals by using a so called proper time representation \cite{schwinger1951green}. After combining them, the denominator factor is shown as:
\beq
\label{eqn_LongAB}
\frac{1}{ab}
=\int_{0}^{1}\dif x\int_{0}^{\infty}\dif\tau~\tau\exp\l[\l(xa+(1-x)b\r)\tau\r]
\eeq
where $a=\kp^2-M^{2}-2neB+i\eps$ and $b=(\kp+\qp)^2-M^{2}-2meB+i\eps$ in our work. $\tau$ is the variable of proper time. Before step on, we normalize the momentum to dimensionless, where $\hat{q}^{2}=q^{2}/(2eB)$, $\hat{M}^{2}=M^{2}/(2eB)$, and so on.

It is common known to shift $k$ to $k-(1-x)q$ to simplify the later calculations. However, we only shift $\hkp$ to $\hkp-(1-x)\hqp$ as usual, but shift transverse momentum $\hat{{k}}_{\perp}$ to $\hat{{k}}_{\perp}-\be\hqt/(\al+\be)$. The reason will show later by the explicit expression of \Eqn{eqn_summation}. Also, the detailed meaning of notation $\al,\be$ are described in \Eqn{eqn_albe}. Now, the tensor structure of vacuum polarization $I^{\mu\nu}$ becomes
\bea
\label{eqn_Imunu}
I^{\mu\nu}&=&
2\Tr\l[\slashed{k}_{\shortparallel}^{1-x}\ga^{\mu}_{\shortparallel}\slashed{k}_{\shortparallel}^{x}\ga^{\nu}\r]
\l({L}_{n}{L}_{m}+{L}_{n-1}{L}_{m-1}\r)
\nn &&
-2\Tr\l[\slashed{k}_{\shortparallel}^{1-x}\ga^{\mu}_{\perp}\slashed{k}_{\shortparallel}^{x}\ga^{\nu}\r]
\l({L}_{n}{L}_{m-1}+{L}_{n-1}{L}_{m}\r)
\nn &&
-4\Tr\l[\slashed{k}_{\shortparallel}^{1-x}\ga^{\mu}\slashed{k}_{\perp}^{\al}\ga^{\nu}\r]
\l({L}_{n}-{L}_{n-1}\r){L}^{1}_{m-1}
\nn &&
-4\Tr\l[\slashed{k}_{\perp}^{\be}\ga^{\mu}\slashed{k}_{\shortparallel}^{x}\ga^{\nu}\r]
{L}^{1}_{n-1}\l({L}_{m}-{L}_{m-1}\r)\nn &&
+16\Tr\l[\slashed{k}_{\perp}^{\be}\ga^{\mu}\slashed{k}_{\perp}^{\al}\ga^{\nu}\r]
{L}^{1}_{n-1}{L}^{1}_{m-1},
\eea
where
\bea
\slashed{k}_{\shortparallel}^{x}=\slashed{k}_{\shortparallel}+x\slashed{q}_{\shortparallel}+M,
&&
\slashed{k}_{\shortparallel}^{1-x}=\slashed{k}_{\shortparallel}-(1-x)\slashed{q}_{\shortparallel}+M,
\nn
\slashed{k}_{\perp}^{\al}=\slashed{k}_{\perp}+\frac{\al}{\al+\be}\slashed{q}_{\perp},&&
\slashed{k}_{\perp}^{\be}=\slashed{k}_{\perp}-\frac{\be}{\al+\be}\slashed{q}_{\perp}.
\eea
The augments of $L^{\al,\be}_{n,m}$ have been abbreviated. Therefore,
\beq\label{eqn_pi_first}
\Pi^{\mu\nu}=
\int\dif\Ga~I^{\mu\nu}\me^{-\l[\hat{M}^{2}-\eta\hqp^{2}+nx+m(1-x)-\hkp^{2}-i\eps\r]\tau},
\eeq
where $\eta=x(1-x)$ and
\bea
\dif\Ga&=&-\mi e^{2}\sum_{n=0}^{\infty}\sum_{m=0}^{\infty}(-1)^{n+m}\int_{0}^{1}\dif x\int_{0}^{\infty}\dif\tau~\tau
\nn &&
\cdot\int\frac{\dif^{2-\eps}\hkp}{(2\pi)^{2}}\int\frac{\dif^{2}\hkt}{(2\pi)^{2}}
\exp\l[2\hat{k}_{\perp}^{2}+2\hat{p}_{\perp}^{2}\r].
\eea

Under the help of generating function of Laguerre polynomials \cite{jeffrey2007table}:
\beq
\sum_{n=0}^{\infty}t^{n}L_{n-i}^{a}(\xi)=\frac{t^{i}}{(1-t)^{a+1}}\exp\l[\frac{-t\xi}{1-t}\r]
\eeq
for $|t|<1$, we are able to evaluate the summation of Landau level by a direct manner. We have
\bea
\label{eqn_summation}
&&\exp\l[2\hat{k}_{\perp}^{2}+2\hat{p}_{\perp}^{2}\r]\sum_{n=0}^{\infty}\sum_{m=0}^{\infty}(-1)^{(n+m)}
z^{\frac{n}{2}}z^{\frac{m}{2}}
\nn &&
\cdot\exp\l[-(nx+m(1-x))\tau\r]L_{n-i}^{a}(-4\hkt\mb{}^{2})L_{m-j}^{b}(-4\hat{p}_{\perp}^{2})
\nn &&
=\frac{t_{1}^{i}t_{2}^{j}}{(1-\tone)^{a+1}(1-\ttwo)^{b+1}}\exp\l[\frac{\al\be}{\al+\be}\hqt^{2}\r]
\nn &&
\cdot\exp\l[(\al+\be)\l(\hkt+\frac{\be}{\al+\be}\hqt\r)^{2}\r]
\eea
where the regulator $z=\me^{-\epst}$ (for $\epst\to 0$) has been included in each summation. The last exponential term explains the unusual shifting of transverse momentum which is early taken in \Eqn{eqn_Imunu}. Here, $\tone=-z^{\frac{1}{2}}\me^{-x\tau}$, $\ttwo=-z^{\frac{1}{2}}\me^{-(1-x)\tau}$, and
\beq
\label{eqn_albe}
\al=\frac{2(1+\tone)}{1-\tone},\qquad \be=\frac{2(1+\ttwo)}{1-\ttwo}.
\eeq

In principle, another term
\bea
J^{\mu\nu}&=&-4\mi\l\{
\Tr\l[\slashed{k}_{\shortparallel}^{1-x}\ga^{1}\ga^{2}\ga^{\mu}\slashed{k}_{\perp}^{\al}\ga^{\nu}\r]
\l({L}_{n}+{L}_{n-1}\r){L}^{1}_{m-1}
\r.\nn &&\l.
+\Tr\l[\slashed{k}_{\perp}^{\be}\ga^{\mu}\slashed{k}_{\shortparallel}^{x}\ga^{1}\ga^{2}\ga^{\nu}\r]
{L}^{1}_{n-1}\l({L}_{m}+{L}_{m-1}\r)\r\}
\eea
should appear in $\Pi^{\mu\nu}$. But it definitively vanishes because of the symmetry between $x\leftrightarrow1-x$. In other words, hold by Ward identity.
\section{Lerch Transcendent Regularization}
\label{sec:zeta}
After summation, it is remarkable that we acquire such a simple term, which takes the form of
\beq
\label{eqn_Pimunu_Full}
\Pi^{\mu\nu}=4\int\dif\Ga_{E}~\frac{\tau \me^{-v_{0}\tau}\mc{I}^{\mu\nu}}{(1-\tone)(1-\ttwo)}\exp\l[h(x,\tau)\hqt^{2}\r],
\eeq
where $v_{0}=\hat{M}^{2}-\eta\hqp^{2}-\mi\eps$ and $h(x,\tau)=\al\be/(\al+\be)$. The integral volume space after performing wick rotation is
\beq
\dif\Ga_{E}=e^{2}\int_{0}^{1}\dif x\int_{0}^{\infty}\dif\tau\int\frac{\dif\Ga_{\kp}}{(2\pi)^{2}}\int\frac{\dif\Ga_{\kt}}{(2\pi)^{2}},
\eeq
where
\bea
\int\dif\Ga_{\kp}&=&\int\dif^{2-\eps}\hkp\exp\l[-\hkp^{2}\tau\r],
\nn
\int\dif\Ga_{\kt}&=&\int\dif^{2}\hkt\exp\l[-(\al+\be)\hkt^{2}\r].
\eea
Note here $k^{0}=\mi k_{E}^{0}$ and $k^2=-k_{E}^{2}$. 

In the appendix \ref{sec:albe}, we detailed explore the value of $h(x,\tau)$. We find that it locates in the interval $[0,1]$ for $x\in[0,1]$ and $\tau\in(0,\infty)$. The feature of Gaussian integrals is that its dominant contribution is coming from the point where the argument of the exponential is stationary \cite{weinberg1995quantum,kuznetsov2003electroweak}. Therefore, from the view of mathematics, when $\hqt^{2}\to 0$, the mainly contributed interval of $h(x,\tau)$ is at its maximum value with $\tau\gg 1$, i.e. $h(x,\tau)=1$. As $\hqt^{2}\to\infty$, $h(x,\tau)$ is at its minimum region $\eta\tau$ to produce stationary points with $\tau\ll 1$. From the view of physics, since the sphere space symmetry is almost not broken for $B\to 0$, $h(x,\tau)$ is closing to the coefficient of $\hqp^2$, i.e., $\eta\tau$. On the other hand, the transverse spaces are decoupling from the whole system as the strength of magnetic field increasing, but the longitudinal space is left. All in all, after considering the magnitude of $\hqt^{2}$, the suitable approximation of $h(x,\tau)$ can be applied as
\begin{equation}
\label{eqn_h}
	h(x,\tau)=\th(\tau)\sech\frac{\qt^{2}}{2eB}+\eta\tau\tanh\frac{\qt^{2}}{2eB}.
\end{equation}
Then, we rewrite \Eqn{eqn_Pimunu_Full} as
\beq
\label{eqn_Pimunu_App}
\Pi^{\mu\nu}=4C\int\dif\Ga_{E}~\frac{\tau \me^{-v\tau}\mc{I}^{\mu\nu}}{(1-\tone)(1-\ttwo)},
\eeq
where $C=\exp\l[\hqt^{2}\sech\hqt^{2}\r]$ and
\begin{equation}
	v=\frac{M^2-\eta \qp^2-\eta\qt^{2}\tanh\frac{\qt^{2}}{2eB}}{2eB}-\mi\eps.
\end{equation}

Following \Eqn{eqn_Imunu}, one obtains the full description of the tensor
\bea
\label{eqn_mcImunu}
\mc{I}^{\mu\nu}
&=&
2\l[-\frac{\eps}{2}\gp^{\mu\nu}\kp^{2}-\eta(2\qp^{\mu}\qp^{\nu}-\gp^{\mu\nu}\qp^{2})+\gp^{\mu\nu}M^{2}\r]
\nn &&
\cdot\l(1+\tone\ttwo\r)
-2\gt^{\mu\nu}\l[\kp^{2}+\eta\qp^{2}+M^{2}\r](\tone+\ttwo)
\nn &&
+4\l(\qp^{\mu}\qt^{\nu}+\qt^{\mu}\qp^{\nu}\r)
\l[\frac{\al(1-x)(1-\tone)\ttwo}{(\al+\be)(1-\ttwo)}+\mathrm{S.T.}\r]
\nn &&
+16\l[(\al+\be)\gp^{\mu\nu}\kt^{2}-\frac{\al\be}{\al+\be}(2\qt^{\mu}\qt^{\nu}-g^{\mu\nu}\qt^{2})\r]
\nn &&
\cdot\frac{\tone\ttwo}{(1-\tone)(1-\ttwo)(\al+\be)},
\eea
where S.T. denotes the related symmetric terms by exchanging $x$ and $1-x$. Both the functions of longitudinal and transverse momentum are at a simple Gaussian form in the integrand. Carrying out the integration, one gets
\beq
\label{eqn_Phi}
\pi^{\mu\nu}=\frac{1}{4\pi^{2}}\int_{0}^{\infty}\dif\tau\frac{\tau^{\frac{\eps}{2}}\me^{-v\tau}}{1-z\me^{-\tau}}
\l\<\mc{I}^{\mu\nu}(\tau^{s})\r\>,
\eeq
where
\beq
\l\<\mc{I}^{\mu\nu}(\tau^{s})\r\>=\tau^{1-\frac{\eps}{2}}(\al+\be)\int\dif\Ga_{\kp}\dif\Ga_{\kt}~\mc{I}^{\mu\nu}(\tau^{s}).
\eeq
The denominator in \Eqn{eqn_Phi} is determined by algebra
\beq
\frac{1}{(1-\tone)(1-\ttwo)(\al+\be)}=\frac{1}{4(1-z\me^{-\tau})}.
\eeq

The result of integrating \Eqn{eqn_Phi} is exactly proportional to $\Ga(s+\frac{\eps}{2})\Phi(z,s+\frac{\eps}{2},v)$ \cite{erdelyi1953higher,jeffrey2007table}, where $\Phi(z,s+\frac{\eps}{2},v)$ is the Lerch transcendent, see the details in appendix \ref{sec:lerch}. Thus, it is straightforward for us to write down the final expression of $\pi^{\mu\nu}=\pi^{\mu\nu}_{a}+\pi^{\mu\nu}_{b}+\pi^{\mu\nu}_{c}$, which are
\bea\label{eqn_pimunu_P}
\pi^{\mu\nu}_{a}&=&
-\frac{\eps}{4}\gp^{\mu\nu}(2eB)\Ga\l(\frac{\eps}{2}\r)\l(\Phi(z,\frac{\eps}{2},v)+\Phi(z,\frac{\eps}{2},v+1)\r)
\nn &&
-\frac{1}{2}\l[\eta\l(2\qp^{\mu}\qp^{\nu}-\gp^{\mu\nu}\qp^{2}\r)-\gp^{\mu\nu}M^{2}\r]
\nn &&
\cdot\l(\Phi(z,1+\frac{\eps}{2},v)+\Phi(z,1+\frac{\eps}{2},v+1)\r)
\nn &&
+\gp^{\mu\nu}(2eB)D_{z}\Phi(z,1+\frac{\eps}{2},v),
\eea
\bea
\pi^{\mu\nu}_{b}&=&
\label{eqn_pimunu_Mix1}
\frac{1}{2}\gt^{\mu\nu}(2eB)D_{z}\Phi(z,1+\frac{\eps}{2},v+x)+\mathrm{S.T.}
\\ &&
\label{eqn_pimunu_Mix2}
+\frac{1}{2}\gt^{\mu\nu}\l(\eta\qp^{2}+M^{2}\r)\Phi(z,1+\frac{\eps}{2},v+x)+\mathrm{S.T.}
\\ &&
\label{eqn_pimunu_Mix3}
-\frac{1}{2}\l(\qp^{\mu}\qt^{\nu}+\qt^{\mu}\qp^{\nu}\r)(1-x)\l(D_{z}\Phi(z,1+\frac{\eps}{2},v-x)
\r. \nn && \l.
-D_{z}\Phi(z,1+\frac{\eps}{2},v+x)\r)+\mr{S.T.},
\eea
\bea\label{eqn_pimunu_T}
\pi^{\mu\nu}_{c}&=&-\l(2\qt^{\mu}\qt^{\nu}-g^{\mu\nu}\qt^{2}\r)\l(\sech\hqt^2 D_{z}\Phi(z,1+\frac{\eps}{2},v)
\r.\nn && \l.
+\eta\tanh\hqt^2 D_{z}\Phi(z,2+\frac{\eps}{2},v)\r).
\eea
Here, not only $\Phi$ is going to divergent for $s$ being non-positive integer at $z\to 1$, but also $\Ga(s+\frac{\eps}{2})$. When $s=0$, the unpleasant divergence of Gamma function is originated from the UV-divergence with $\kp^{2}$ in the numerator. Loosely speaking, the dimension regularization in the longitudinal space should reduce such quadratic divergence to a logarithmic one, and then $s$ has to be greater than zero while evaluating \Eqn{eqn_Phi}. To cure the divergent $\Ga(\frac{\eps}{2})$, we employ the formula
\beq
\frac{1}{\tau}\simeq\frac{z\me^{-\tau}}{1-z\me^{-\tau}}\bigg|_{\tau\to 0, z\to 1}
\eeq
to increase the power of $\tau$ and therefore get the associated term in \Eqn{eqn_pimunu_Mix1}. We also adopt the approximate form of $h(x,\tau)$ to get \Eqn{eqn_pimunu_T}.

We separate $\Pi^{\mu\nu}$ into three parts, in the form of
\beq
\Pi^{\mu\nu}=\frac{e^2 C}{4\pi^{2}}\int_{0}^{1}\pi^{\mu\nu}~\dif x =\frac{e^2 C}{4\pi^{2}}P^{\mu\nu}_{i}(q^2)\pi_{i},
\eeq
where $\pi_{i}=\pi_{\shortparallel},\pi_{\perp},\pi_{m}$ correspond to longitudinal, transverse and mixture scalar function, respectively. The projecting operators are
\beq
P_{\shortparallel}^{\mu\nu}(q^{2})=\qp^{\mu}\qp^{\nu}-\gp^{\mu\nu}\qp^{2}
,~P_{\perp}^{\mu\nu}(q^{2})=\qt^{\mu}\qt^{\nu}-\gt^{\mu\nu}\qt^{2}
\eeq
and
\beq
\label{eqn_tensor_mix}
P_{m}^{\mu\nu}(q^{2})=q^{\mu}q^{\nu}-g^{\mu\nu}q^{2}-P_{\shortparallel}^{\mu\nu}(q^{2})-P_{\perp}^{\mu\nu}(q^{2}).
\eeq
Led by the explicit form of \eqn{eqn_pimunu_P}, we get
\bea
\label{eqn_pi_p}
\pi_{\shortparallel}&=&-\int_{0}^{1}\eta\l[\Phi(z,1,v)+\Phi(z,1,v+1)\r]\dif x
\nn
&=&\int_{0}^{1}\eta\l[\psi(v)+\psi(v+1)\r]\dif x
\eea
where the logarithmic divergence has been removed, see \Eqn{eqn_Phi_z0}. $\psi$ is di-gamma function, which is the logarithmic derivative of the Gamma function, $\psi(v)=\p_{v}\ln\Ga(v)$. The detailed limiting behaviors of function $\Phi$ are discussed in appendix \ref{sec:lerch}. Meanwhile, from the summation representative of Lerch transcendent \cite{erdelyi1953higher,jeffrey2007table}, we obtain
\beq
D_{z}\Phi(z,s,v)=\Phi(z,s-1,v+1)-v\Phi(z,s,v+1).
\eeq
Hence, the transverse scalar self-energy function can be derived as:
\bea
\label{eqn_pi_t}
\pi_{\perp}&=&
2\int_{0}^{1}\l(\eta v\tanh\hqt^2\zeta(2,v+1)-\sech\hqt^2\zeta(0,v+1)
\r. \nn && \l.
+\psi(v+1)\l(\eta\tanh\hqt^2-v\sech\hqt^2\r)\r)\dif x
\eea
via formula \eqn{eqn_pimunu_T}. The mixture scalar function is described as
\beq
\label{eqn_pi_m}
\pi_{m}=\int_{0}^{1}\eta\l(\psi(v+x)+\psi(v+1-x)\r)\dif x.
\eeq
The full tensor structure of $P_{m}^{\mu\nu}$ is combined by several components, including $\pi^{\mu\nu}_{b}$ and a partial of \Eqn{eqn_pimunu_P} and \Eqn{eqn_pimunu_T}. Their associated scalar functions are closed but not exactly equivalent in our current calculation, shown by \eqn{eqn_pimunu_Mix1}, \eqn{eqn_pimunu_Mix2}, \eqn{eqn_pimunu_Mix3}, and so on. As a gauge invariant tensor, such slight differences should not exist. But, they are inevitable since we have to simplify $h(x,\tau)$ to access a final answer. Here, we pick up the typical and dominant parts among them as our results of $\pi_{m}$, given in \Eqn{eqn_pi_m}.
\section{Discussion of Results}
\label{sec:res}
Di-Gamma function $\psi(v)=-1/v+\psi(v+1)$ has a simple pole at $v=0$ with residue $-1$ \cite{erdelyi1953higher,jeffrey2007table}. It leads that $\Pi^{\mu\nu}$ is not a purely real valued function but contains imaginary part when the variable of $\psi$ becoming negative. Since the maximum value of $\eta=x(1-x)$ is at the point $x=1/2$, the threshold condition is $\qp^2=4M^2$ for strong magnetic fields, where transverse momenta have been decoupled. The reason that the threshold condition is not sensitive to the magnitude of magnetic fields is because the virtual pair gains energy from the external momentum of photon and becomes real via photon decay. One should distinguish it with the pair production in strong electric fields via Schwinger effect, i.e. virtual pair gains energy from the external electric field \cite{dunne2005heisenberg}.

{{\textbf{Strong Magnetic Field Limit}}}~For $eB\gg \qt^2$, one has $C=\exp\l[\hqt^2\r]$ and $v=\hat{M}^{2}-\eta\hqp^{2}$. Plus, both $\hqp^2$ and $\hat{M}^2$ are not greater than 1 for $eB\gg \qp^{2}$ and $eB\gg M^2$. The maximal value of $\lfloor \hqp^{2}/4-\hat{M}^2\rfloor$ (floor function) is equal to zero. Therefore, among three scalar functions, only $\pi_{\shortparallel}$ contains imaginary part in the small distance regime where $\qp^2>4M^2$, which is
\beq
\Im \pi_{\shortparallel}=\frac{-4\pi eBM^2}{\qp^2\sqrt{\De_{\shortparallel}}},
\eeq
where $\De_{\shortparallel}={\qp^4-4\qp^2M^2}$. Since here we have $|v|\ll 1$, $\psi$ can be expanded as \cite{jeffrey2007table}:
\beq
\psi(v+1)=-\ga+v\zeta_{2}+\mc{O}(v^2).
\eeq
$\ga\approx 0.577216$ is the Euler-Gamma constant and $\zeta_{2}=\pi^{2}/6$ . It results in
\beq
\label{eqn_pi_p_strong}
\Re \pi_{\shortparallel}=
\left\{
  \begin{array}{ll}
    \pi_{0}-\frac{8eBM^2\mathrm{arctanh}\frac{\qp^2}{\sqrt{\De_{\shortparallel}}}}{\qp^2\sqrt{\De_{\shortparallel}}}, & \hbox{\small{$\qp^2> 4M^2$};} \\
    \pi_{0}+\frac{8eBM^2\arctan\frac{\qp^2}{\sqrt{-\De_{\shortparallel}}}}{\qp^2\sqrt{-\De_{\shortparallel}}}, & \hbox{\small{$\qp^2\leq 4M^2$},}
  \end{array}
\right.
\eeq
where $\pi_{0}=-2eB/\qp^2-\ga/3$. Furthermore, one deduces that $\Re\pi_{\shortparallel}=\pi_{0}$ for $\qp^2\gg M^2$ and $\pi_{\shortparallel}=eB/(2M^2)-\ga/3$ for $\qp^2\ll M^2$. $\pi_{\perp}$ and $\pi_{m}$ are purely real valued functions at this limit. One gets that
\beq
\pi_{m}=-4\ln A,~~\pi_{\perp}=\mc{O}\l(\frac{\La^2}{eB}\r),
\eeq
where $A\approx 1.28243$ is the Glaisher-Kinkelin constant and $\La^{2}\sim M^2$ and/or $q^2$. Evidently, one gets $\pi_{\shortparallel}\gg \pi_{m}\gg \pi_{\perp}$ at strong $B$ limit, which demonstrates the decoupling of transverse spaces.

{\textbf{Medium Regime}}~Di-Gamma function $\psi$ is meromorphic with simple poles at not only zero point but also $v=-1,-2,...$ with residues $-1$. Let $\tilde{q}^2=\qp^{2}+\qt^{2}\tanh\hqt^{2}$, one finds that only $\pi_{\shortparallel}$ becomes complex while $\tilde{q}^2>4M^2$ but $\tilde{q}^2\leq 2eB$. The imaginary component of $\pi_{m}$ arise while $\tilde{q}^2>2M^2+2eB+2\sqrt{M^4+2eBM^2}$. All three scalar functions are characterized by an imaginary part while $\tilde{q}^2>4(M^2+2eB)$. Thus, if $n=\lfloor \tilde{q}^2/(8eB)-\hat{M}^2\rfloor$ is a finite positive integer, $\psi$ can be transformed through the following series \cite{jeffrey2007table}
\beq
\psi(v)=\psi(v+n+1)-\sum_{k=0}^{n}\frac{1}{v+k}.
\eeq
Then, one has
\beq
\Im {\pi}_{\shortparallel}=\sum_{k=0}^{n}\frac{-8\pi eBM^2_{k}}{\tilde{q}^2\sqrt{\De_{k}}}+\frac{4\pi eBM^{2}}{\tilde{q}^2\sqrt{\De_{0}}},
\eeq
where $M^2_{k}=M^2+2keB$ and $\De_{k}=\tilde{q}^4-4\tilde{q}^2M^2_{k}$. The physical explanation of $n$ is the highest Landau level which possibly occupied by the created real fermion-anti-fermion pair. The transverse function takes the form of:
\beq
\Im \pi_{\perp}=\sum_{k=1}^{n}\frac{-8\pi eBM^2_{k}}{\tilde{q}^2\sqrt{\De_{k}}}\tanh\hqt^2+\frac{8\pi keB}{\sqrt{\De_{k}}}\sech\hqt^2.
\eeq
Besides, let $n=\lfloor \tilde{q}^{2}/(8eB)+eB/(2\tilde{q}^2)-\hat{M}^2-1/2\rfloor$, one has
\beq
\Im \pi_{m}=\sum_{k=0}^{n}\frac{-8\pi eB\l(\tilde{q}^2M^{2}_{k+\frac{1}{2}}-2e^2B^2\r)}{\tilde{q}^4\sqrt{\Theta_{k}}},
\eeq
where $\Theta_{k}=\tilde{q}^4+4e^2B^2-4\tilde{q}^2M^{2}_{k+\frac{1}{2}}$.

Integrating analytically with respect to $x$ is not a plain task. However, the numerical evaluation of a complete result of $\Pi^{\mu\nu}$ in any kinematic regime and any strength of $B$-field is not computationally expensive. We leave it in the future work. An interesting algebra given here is that
\beq
\label{eqn_pimunu_weak_B_Int}
\int_{0}^{n}\frac{2eBM^2_{k}}{q^2\sqrt{\De_{k}}}\dif k=\frac{1}{12}\sqrt{1-\frac{4M^2}{q^2}}\l(1+\frac{2M^2}{q^2}\r).
\eeq

{\textbf{Weak Magnetic Field Limit}}~The fermion propagator at a constant magnetic field reduces to a free one if $B\to 0$. It indicates that the three scaler functions have to become uniform and close to the usual results without $B$-fields. For $eB\ll q^2$ and $eB\ll M^2$, $C=1$ and $\tilde{q}^2=q^2$. Especially, $v$ is possible going to either $+\infty$ or $-\infty$. At these regimes, $\psi(v)$ is logarithmically divergent \cite{erdelyi1953higher}. A counter term of course has to be introduced to regularize results. According to the properties of $\psi$, one has \cite{erdelyi1953higher}
\beq
\psi(v)=\frac{1}{m}\sum_{k=0}^{m-1}\psi\l(\frac{v+k}{m}\r)+\ln m.
\eeq
Let $m=\lceil\hat{M}^{2}\rceil$ (ceiling function). The regularized $\hat{\psi}(v)$ is defined by $\hat{\psi}(v)=\psi(v)-\ln m$. The counter term $\ln m$ is zero at strong magnetic field limit, i.e., $m=1$ for $M^2\ll eB$. Another important property of counter term is that it is a purely real valued function and should not affect the imaginary part of $\Pi^{\mu\nu}$. Through $\Im\psi(v)=\Im\ln v$ at $v\to-\infty$, one can easily derive the imaginary part of the complex scalar functions for $q^2\gg M^{2}$:
\beq
\label{eqn_pimunu_weak_B_Im}
\Im \pi_{\shortparallel}=\pi_{\mathbb{I}}+\frac{4\pi eBM^2}{q^2\sqrt{\De}};~~\Im \pi_{\perp}=\pi_{\mathbb{I}}+\frac{8\pi eBM^2}{q^2\sqrt{\De}}.
\eeq
Note that here 
\beq
\pi_{\mathbb{I}}=\frac{-\pi}{3}\sqrt{1-\frac{4M^2}{q^2}}\l(1+\frac{2M^2}{q^2}\r).
\eeq
The results of \Eqn{eqn_pimunu_weak_B_Im} are exactly the same as the ones if performed via early formula \eqn{eqn_pimunu_weak_B_Int}. Besides,
\beq
\label{eqn_pimunu_weak_B_Im_Mix}
\Im \pi_{m}=\frac{-\pi}{3}\frac{\sqrt{\Th_{0}}}{q^2}\l(1+\frac{4e^2B^2}{q^4}+\frac{2M^2+2eB}{q^2}\r).
\eeq
Obviously, when $|v|\to\infty$, the differences among three scalar functions are at the order of $eB/\La^{2}$, examined by above results. Thus, we only provide the real part of the scalar function $\pi_{\shortparallel}$ at the weak magnetic field limit, which is
\beq
\label{eqn_pi_weak_Mass}
\Re\hat{\pi}_{\shortparallel}=\frac{1}{m}\l(\sum_{k=0}^{m-1}+\sum_{k=1}^{m}\r)\int_{0}^{1}\psi\l(\frac{M^2_{k}-\eta q^2}{M^2}\r) \dif x.
\eeq
The limiting behaviors are in the forms of:
\beq
\label{eqn_pi_full_weak}
\Re\hat{\pi}_{\shortparallel}|_{q^2\gg M^2}=\frac{1}{3}\l(\ln\frac{q^2}{M^2}-\frac{5}{3}\r)+\frac{2M^2}{q^2};
\eeq
\beq
\hat{\pi}_{\shortparallel}|_{q^2\ll M^2}=\frac{-\ga}{3}+\frac{\zeta_{2}}{3}\l(\frac{1}{2}-\frac{q^2}{5M^2}\r).
\eeq
\section{Conclusions and Outlooks}
\label{sec:con}
In this paper, we evaluated the photon vacuum polarization tensor in the presence of a homogeneous, purely magnetic field in a systematic framework. The quadratic UV divergence of longitudinal momentum is cured by dimension regularization procedures. We developed the zeta function technique to deal with the infinite series and then subtract the finite results. Therefore, beyond LLL approximation, we obtained a full description of vacuum polarization tensor in response to all the Landau levels at any field strength of $B$ for the first time. The final answers are characterized by the di-Gamma function $\psi(v)$ and zeta functions. Remarkably, we caught an intuitive understanding of the imaginary part of the photon polarization beyond the threshold $\tilde{q}^2=4M^2$ in our context. That is, the singular behavior of $\psi(v)$ near each non-positive integer $-k$ describes the probability which a photon decays into a real fermion-anti-fermion pair at the given Landau level state $k$. The allowed quantum states are determined by the magnitude of $n=\lfloor\tilde{q}^{2}/(8eB)-\hat{M}^2\rfloor$, with $k=0,1,...,n$. Such kind of pair creation process was observed by Hattori and Itakura \cite{hattori2013vacuum}. Their work, and other similar ones \cite{karbstein2013photon,ishikawa2013numerical}, used an alternative method for calculating the magnetized vacuum polarization tensor, which is based on an integration of double proper times.

Below the threshold, $\tilde{q}^{2}<4M^{2}$, the vacuum polarization tensor is a purely real valued function. In the specified physical parameter regime, strong field limit, we analog our answers with the early ones under LLL approximation. We confirmed that LLL approximation is valid in the region where $M^2\gg \qp^2$ but not correct in the region $\qp^2\gg M^2$ \cite{calucci1994nonlogarithmic,gusynin1996dimensional}, where the imaginary part has been missed. We then focused on another interesting kinematics regime, weak field limit. We found out the correct counter term and demonstrated that $\Pi^{\mu\nu}$ is limited to the usual vacuum polarization at $B\to 0$.

As showed by above sections, the integration with respect to the proper time is carried out by Lerch transcendent  $\Phi(z,s,v)$ straightforwardly. It also should be realized as a performance of double summation with respect to two charged virtual particles. Our current work for the first time brings a complete and surprising expression to the photon vacuum polarization at a constant external magnetic field. Our approach is based upon a consistent applying of Lerch transcendent  regularization. The final description is rendered in a simple language of Lerch transcendent  or its $z$-derivation. The whole procedure is elegant and mathematically rigorous. Therefore, it allows us to enlarge the scope of this technique. In other words, Lerch transcendent  can be served as a fundamental tool to sum the infinite series with respect to the Landau levels. It has opened a new window to investigate a wide range of phenomena in all physical regimes. Especially, we expect our results will play an important role in studying the observations in the heavy ion collisions at RHIC and LHC experiments. 

In the next paper, we will extend our studies to finite temperatures. In particular, we will compute the gap equation via the complete photon vacuum polarization tensor. It had been studied under the assumption which the dynamical mass of fermion is determined within the LLL approximation. One should recalculate this essential ingredient based on our presented results.
\appendix
\section{Lerch Transcendent}
\label{sec:lerch}
The integral representative of Lerch transcendent $\Phi(z,s,v)$ is
\beq
\label{eqn_Lerch_int}
\Phi(z,s,v)=\frac{1}{\Ga(s)}\int_{0}^{\infty}\frac{\tau^{s-1}\exp(-v\tau)}{1-z\me^{-\tau}}~\dif\tau,
\eeq
for $|z|\leq1, z\neq1, \Re s>0$ or $z=1, \Re s>1$. While as the summation representative is given by
\beq
\label{eqn_Lerch_sum}
\Phi(z,s,v)=\sum_{n=0}^{\infty}z^{n}(n+v)^{-s}.
\eeq
for $|z|<1, v\neq 0,-1,-2,...$  \cite{erdelyi1953higher,jeffrey2007table}. Rendered by the integral representative, its derivation with respect to $z$ is:
\beq
D_{z}\Phi(z,s,v)=\frac{1}{\Ga(s)}\int_{0}^{\infty}\frac{z\me^{-\tau}\tau^{s-1}\exp(-v\tau)}{(1-z\me^{-\tau})^{2}}~\dif\tau,
\eeq
where $D_{z}=z\p_{z}$.

Besides, $\Phi(1,s,v)=\zeta(s,v)$, where $\zeta$ is the Riemann's Zeta Function. $\zeta(s,v)$ has a meromorphic continuation in the $s$ plane, its only singularity in $\mb{C}$ being a simple pole at $s = 1$ with residue 1. Therefore, we get the limiting behaviors of the Lerch transcendent as below \cite{erdelyi1953higher}:
\beq
\label{eqn_Phi_z0}
\lim_{z\to 1}\Phi(z,1,v)=-\log(1-z)-\psi(v)
\eeq
for s=1. When $\Re(s)<1$,
\beq
\label{eqn_Phi_zs}
\lim_{z\to 1}\Phi(z,s,v)=\frac{\Ga(1-s)}{(1-z)^{1-s}}+\zeta(s,v).
\eeq
Another important identity of $\Phi$ is
\beq
\Phi(z,s,v)=z\Phi(z,s,v+1)+\frac{1}{v^{s}}.
\eeq

\section{$h(x,\tau)$}
\label{sec:albe}
Remind you that $\tone=-z^{\frac{1}{2}}\me^{-x\tau}$, $\ttwo=-z^{\frac{1}{2}}\me^{-(1-x)\tau}$ and
\begin{equation}
\label{eqn_albe_exp}
	h(x,\tau)=\frac{\al\be}{\al+\be}=\frac{1+\tone+\ttwo+\tone\ttwo}{1-\tone\ttwo}.
\end{equation}
Because the $n$-th Bernoulli polynomials $B_{n}(x)$ represent the coefficients of $\tau^{n-1}/n!$ in the expansion of the generating function \cite{jeffrey2007table}:
\beq
\frac{\me^{x\tau}}{\me^{\tau}-1}=\sum_{n=0}^{\infty}B_{n}(x)\frac{\tau^{n-1}}{n!},
\eeq
\Eqn{eqn_albe_exp} is deduced to
\bea
\label{eqn_B_eta}
h(x,\tau)\big|_{z\to 1}&=&2\sum_{m=1}^{\infty}(B_{2m}(0)-B_{2m}(x))\frac{\tau^{2m-1}}{(2m)!}
\nn
&=& \eta\tau+\mc{O}(\tau^{3}),
\eea
for $\tau <1$. However, for $\tau>1$, we have to use another formula, which
\bea
h(x,\tau)\big|_{z\to 1}&\leq& \frac{1-2\sqrt{\tone\ttwo}+\tone\ttwo}{1-\tone\ttwo}=\tanh\frac{\tau}{4}.
\eea
Generally, $\tanh\frac{\tau}{4}$ can be replaced by step fucntion $\th(\tau)$ and its maximum value is equal to 1.


\vspace{1.5em}
{\bf\it Acknowledgement.---}
We thank K. Hattori for pointing out a severe mistake. We also thank I. Shovkovy for useful discussions. This work is supported by the NSFC under Grant No. 11275213, DFG and NSFC (CRC 110), CAS key project KJCX2-EW-N01 and Youth Innovation Promotion Association of CAS. This Project of J. C. is supported by China Postdoctoral Science Foundation Grant No. 2013M530732.


\end{document}